\documentclass{IEEEtran}
\usepackage{graphicx}
\usepackage{subcaption}
\usepackage{amsmath}
\usepackage{makecell}
\usepackage{amssymb} 
\begin{document}
\title{Skipped Adjacency Pulse Width Modulation:\\ Zero Voltage Switching over Full Duty Cycle Range for Hybrid Flying Capacitor Multi-Level Converters without Dynamic Level Changing}

\author{Inhwi~Hwang, {inhwi@umich.edu}}

\markboth{Power Electronics and Control Report, 2024}%
{Shell \MakeLowercase{\textit{et al.}}: Bare Demo of IEEEtran.cls for IEEE Journals}

\maketitle

\begin{abstract}
This paper proposes a method to achieve zero voltage switching (ZVS) across the full duty cycle range in hybrid flying capacitor multilevel (FCML) converters, eliminating the need for dynamic level changing and active re-balancing. Utilizing skipped adjacency pulse width modulation (SAPWM), this approach avoids the nearest pole voltage level, thereby increasing volt-seconds within specific duty cycle range. The method uses a modified PWM scheme, which preserves effective pole voltage by changing duty reference and employing digital logic processing. Simulation results verify the proposed method achieving full-range ZVS. This SAPWM technique is compatible with hybrid FCML converters with various levels, offering enhanced efficiency and reduced switching losses.
\end{abstract}


\IEEEpeerreviewmaketitle

\section{Background}
\begin{figure}[t]
    \centering
    \includegraphics[width=1\linewidth]{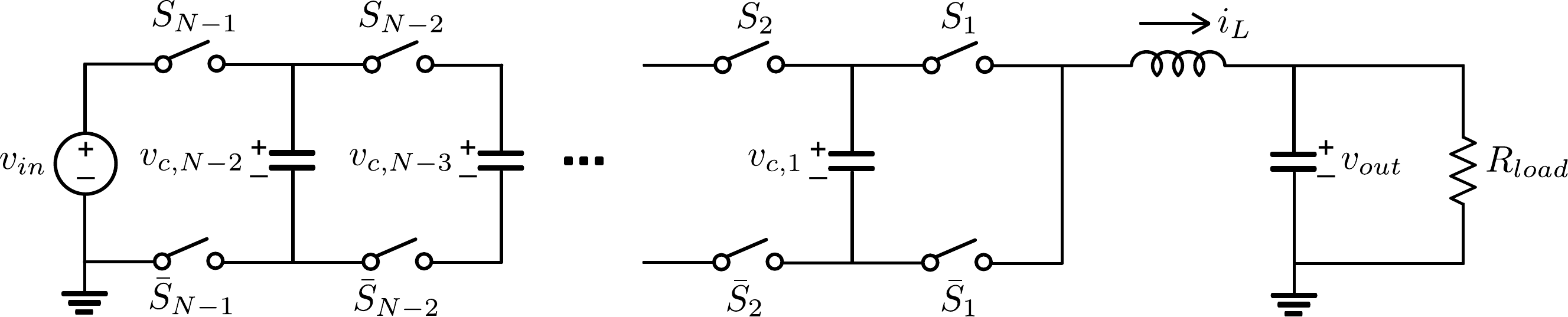} 
    \caption{Circuit diagram of hybrid FCML topology.}
    \label{fig:FCML} 
\end{figure}

\IEEEPARstart{H}{ybrid} flying capacitor multilevel
(FCML) converter as shown in Fig. \ref{fig:FCML}, which effectively reduces inductor volt-seconds and increases effective switching frequency, has gained attention for enhancing power density in power converters \cite{254717}. By increasing the switching frequency, power density can be improved, allowing for reduced volumes of energy storage components such as inductors and capacitors \cite{10261415}. However, higher switching frequencies also lead to increased switching losses, which degrade power efficiency. To mitigate the switching loss, zero voltage switching (ZVS) techniques can be applied \cite{9351687, 7961198}. A common approach is the use of variable switching frequencies to achieve ZVS, as widely investigated to date \cite{10684601, 10548, 8341162}.

In FCML topologies, however, certain duty ranges require exceptionally low switching frequencies to achieve ZVS, a limitation that arises due to the extremely low volt-seconds. To address this, recent work has introduced dynamic level changing to increase volt-seconds \cite{8460086, 9864026}. The switching frequency for ZVS across the duty cycle range is chosen to prevent excessively low switching frequencies in 4-level and 5-level FCML converters. To avoid extremely low switching frequencies, the switching level is changed according to the duty cycle, allowing the system to achieve an appropriate switching frequency for ZVS.
This enables almost full-range ZVS, making suitable for applications requiring variable duty at steady-state, such as AC-DC and DC-AC conversion \cite{10613943, 10221009}.

Nevertheless, the dynamic level changing approach presents the following limitations:

\begin{center} 
\textbf{(1)} 
Dynamic level changing inherently requires active re-balancing during level transitions, which increases the control complexity.
\end{center} 

\begin{center} 
\textbf{(2)} 
Active re-balancing during dynamic level changing could not achieve ZVS instantaneously for rapidly and continuously changing AC voltage input/output, due to the settling time required for active balancing.
\end{center} 
\textit{Note: In literature \cite{9864026}, active balancing requires approximately 80 $\mu$s per operation. In grid-tied conversion applications with a 60 Hz line frequency, a total of 12 rebalancing events are required, summing to around 1 ms, 8\% of the 13.3 ms per a line voltage cycle.}

\begin{center} 
\textbf{(3)} 
To implement dynamic level changing, more switches are required than the optimize number required to fully utilize the voltage rating of switches effectively \cite{7556759}.
\end{center} 
\textit{Note: For a dynamic level changing with 4- and 5- levels, the switch voltage rating must be designed to meet the requirements of the 4-level configuration.}

\begin{center} 
\textbf{(4)} 
Additional conduction loss is induced due to the additional conduction path resulting from the increased levels in the FCML.
\end{center} 

This study addresses the identified research gap by proposing a new method to achieve full-range ZVS. The approach eliminates the need for dynamic level changing, active re-balancing, and additional switch sets. Instead, it relies solely on modified switching techniques. The implementation is feasible by maintaining the phase-shifted pulse width modulation (PSPWM) carrier of the micro-controller unit (MCU) and adding a simple digital circuit, making the proposed method adaptable to various FCML levels.

The following content is organized into three sections: Chapter II introduces conventional ZVS methods using PSPWM and variable switching frequency, highlighting the limitations of PSPWM for ZVS operation. Chapter III presents the proposed PWM technique for achieving ZVS across the full duty range. Chapter IV demonstrates the applicability of the proposed method through simulation results. Chapter V includes future work.

\section{Zero Voltage Switching with Variable Switching Frequency}

Sampling the inductor current at the PWM carrier's peak or valley, assuming negligible series resistance, accurately captures the average current over the switching period. This approach is widely adopted, as it inherently filters out switching ripple effects on the controller. This sampling technique is also employed in FCML control.

\begin{figure}[t]
    \centering
    \includegraphics[width=0.9\linewidth]{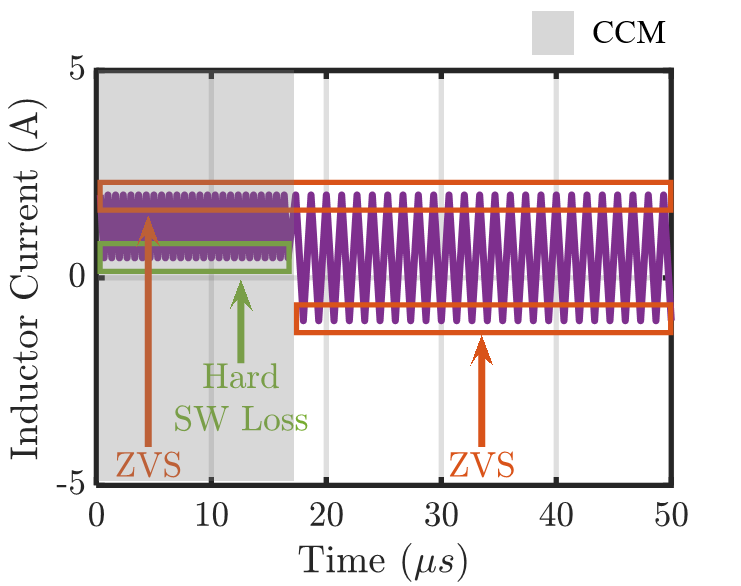} 
    \caption{Inductor current waveform indicating ZVS and hard switching under two different switching frequencies.}
    \label{fig:ZVS} 
\end{figure}

The inductor current is controlled by applying an averaged voltage through PWM. The switching ripple generated by PWM arises from the integration of the instantaneous pole voltage ($v_{sw}$). In continuous current mode (CCM) control, this ripple induces hard-switching in roughly half of the switching events, dissipating the energy stored in the switch’s output capacitance and contributing to losses as depicted in Fig. \ref{fig:ZVS}.

\begin{figure}[t]
    \centering
    \includegraphics[width=0.9\linewidth]{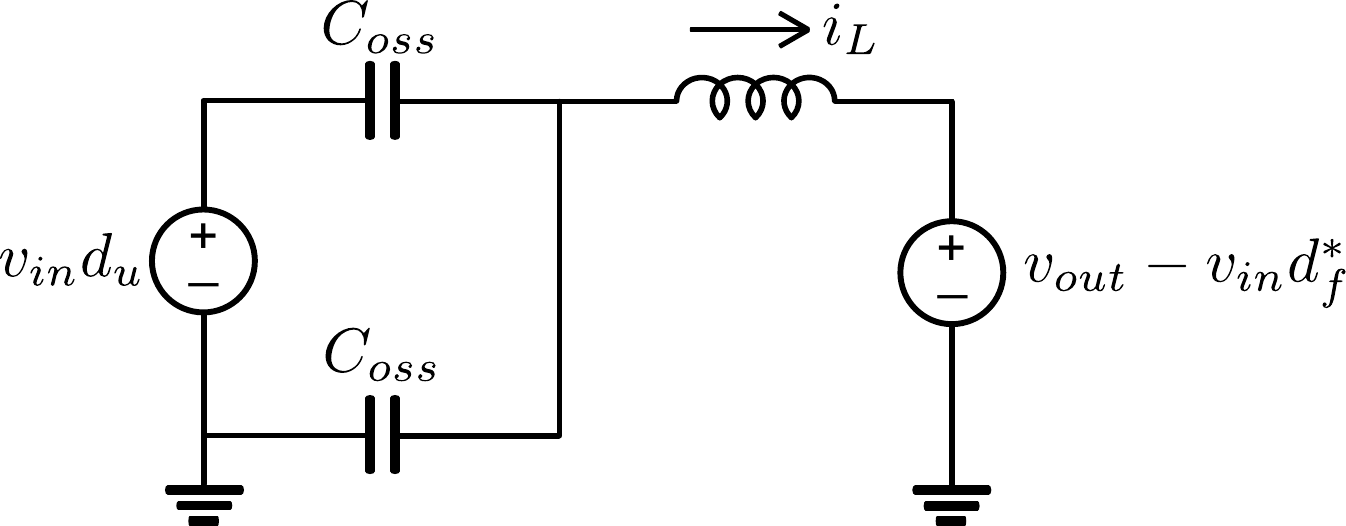} 
    \caption{Equivalent LC resonant circuit between the switch output capacitors and inductor under PSPWM during dead-time.}
    \label{fig:basic_res} 
\end{figure}

Meanwhile, to prevent short circuit of flying capacitors and DC-link caused by non-ideal delays such as PWM signal delay and turn-on/off delay, a dead-time is applied before changing switching states. During this dead-time, both complementary switches are off, allowing resonance between the output capacitor of switches ($C_{oss}$) and inductor ($L$) as shown in Fig. \ref{fig:basic_res}, which charges and discharges the output capacitors of switches. When one of the switch output capacitor is fully discharged, current flows through the anti-parallel diode of the switch. At this point, turning on the switch with the fully discharged output capacitor enables zero voltage switching, allowing for mitigated loss in switching.

The duty cycle reference ($d^{*}$) and following switching state sequences ($S_{k,in}$) are determined by the current controller, so ZVS requires setting the sign of inductor current appropriately at each switching event. When the upper switch turns on, the current should be negative to allow conduction through the upper diode. Conversely, when the lower switch turns on, the current should be positive to conduct through the lower diode. Additionally, current decreases when the lower switch is on and increases when the upper switch is on. Thus, the current at the peak and valley where switching occurs should be ensured to have opposite signs for ZVS as shown in Fig. \ref{fig:ZVS}.
\begin{figure}[t]
    \centering
    \includegraphics[width=0.9\linewidth]{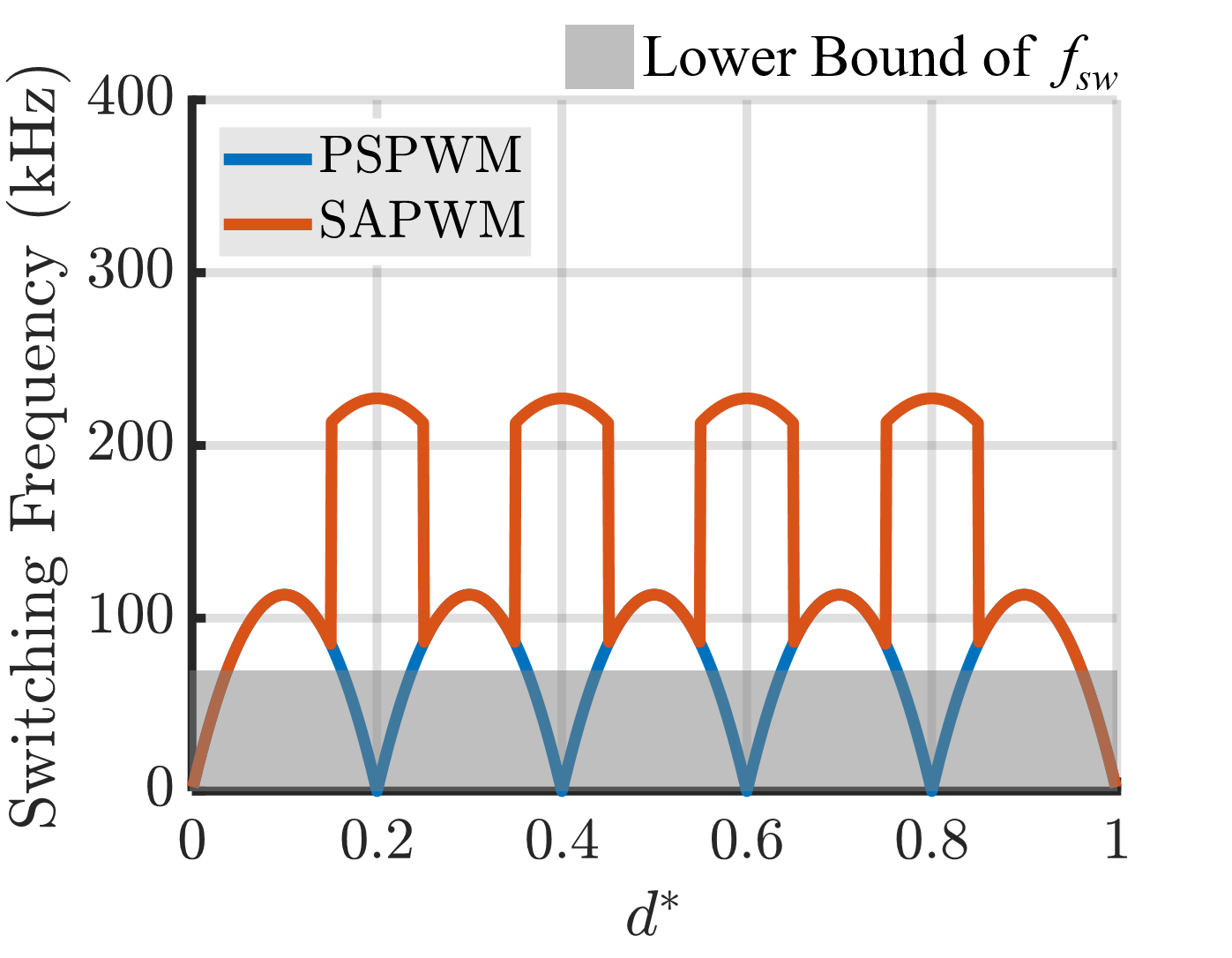} 
    \caption{Variable frequency profile for ZVS under PSPWM and SAPWM, given parameters: $I_{ZVS} = 1$ A, $|i_L| = 3$ A, $L = 4.4$ $\mu$H, $V_{in} = 400$ V, and $N = 6$. The lower bound of switching frequency is 70 kHz.}
    \label{fig:fsw} 
\end{figure}
\begin{figure}[t]
    \centering
    \includegraphics[width=0.9\linewidth]{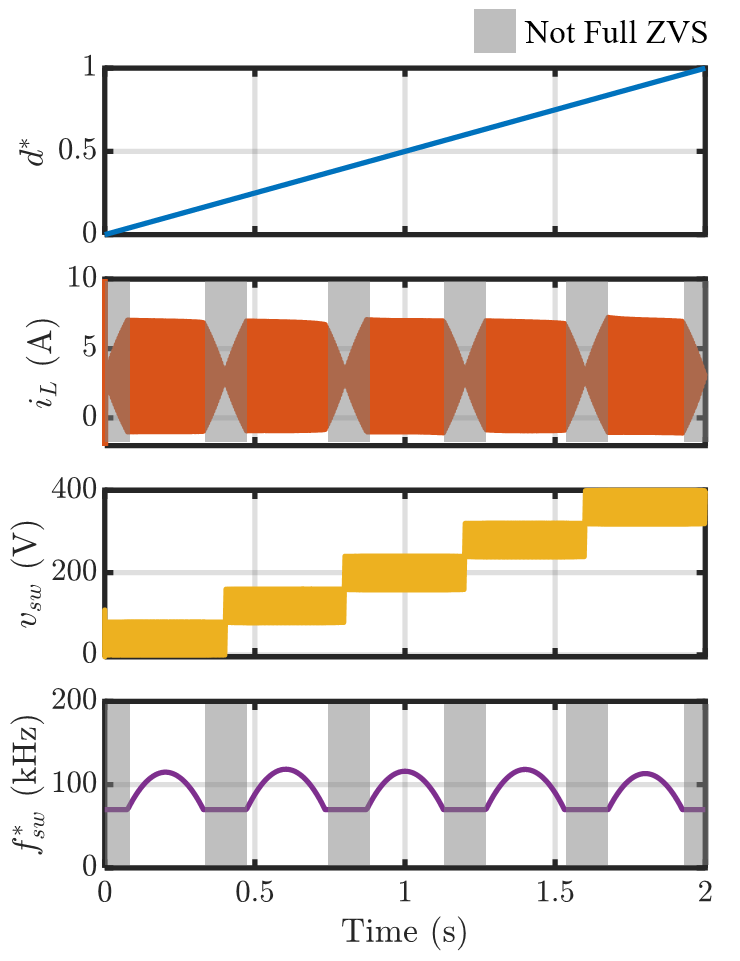} 
    \caption{Current control result with PSPWM and variable switching frequency, including a lower bound of 70 kHz for switching frequency. }
    \label{fig:PSPWM_sim} 
\end{figure}

During dead-time ($d_{t}$), the stored energy in inductor must remain sufficiently higher than the output capacitor’s energy to fully discharge the output capacitor. Thus, peak and valley currents should be set large enough to ensure this condition. However, if these currents are too high, the increased RMS current may lead to higher conduction losses. Moreover, extending dead-time can also be considered, but this may increase conduction loss due to diode conduction. Therefore, they should be set with adequate margin but not excessively high. Assuming the inductor energy is sufficiently large compared to the stored energy of the switch’s output capacitor, full-range ZVS can be achieved under the following condition:
\begin{equation}
\frac{2{{C}_{oss}}{{d}_{u}}{{v}_{in}}}{{{I}_{ZVS}}}\le {{t}_{d}}
\label{ineq}
\end{equation}, where $I_{ZVS}$ is the current magnitude for ZVS.

When the average power in a single switching cycle flows from input to output, the sampled inductor current ($i_{L}$) is positive, ensuring that the peak current naturally aligns with the ZVS current polarity. However, the valley current may not always satisfy the ZVS condition, depending on the switching frequency. In this case, the switching frequency should be adjusted so that the valley current meets the ZVS condition. Conversely, when power flows from output to input, $i_{L}$ is negative, and the switching frequency ($f_{sw}$) must be modified so that the peak current satisfies the ZVS condition. The switching frequency required to achieve ZVS with peak and valley currents under PSPWM is determined as follows:
\begin{equation}
{{f}_{sw}}^{*}=\frac{\left\{ {{v}_{in}}\left( {{d}_{f}}^{*}+{{d}_{u}} \right)-{{v}_{out}} \right\}\left( d^{*}-{{d}_{f}}^{*} \right)}{2L\left( \left| {{i}_{L}} \right|+{{I}_{ZVS}} \right)}
\label{freq_ZVS}
\end{equation}
\begin{equation}
{{d}_{f}}^{*}=\frac{floor\left( \left( N-1 \right){{d}^{*}} \right)}{N-1} 
\end{equation}
\begin{equation}
{{d}_{u}}=\frac{1}{N-1}
\end{equation}
The switching frequency reference ($f^{*}_{sw}$) is set so that the smaller of the peak or valley current reaches $\pm$${{I}_{ZVS}}$, respectively (${{I}_{ZVS}}$ $>$ 0).

However, the multi-level operation of the FCML leads certain duty cycles to naturally produce small or zero current ripple. This limits the ability to achieve ZVS, as it necessitates extremely low switching frequencies, resulting in prolonged charging and discharging periods for the flying capacitors. Thus, ZVS operation with variable switching frequency, as derived from (\ref{freq_ZVS}), is limited by a lower bound. This constraint restricts full-range operation for ZVS in variable switching frequency operation, allowing ZVS only at certain operating points as shown in Fig. \ref{fig:PSPWM_sim}.
\section{Achieving ZVS Across the Full Duty Cycle Range with SAPWM}
The adjacent duty cycle reference, $d_{r}^{*}$, is defined with $d^{*}$ as follows:
\begin{equation}{{d}_{r}}^{*}=\frac{round\left( \left( N-1 \right){{d}^{*}} \right)}{N-1}
\end{equation}
As $d^{*}$ approaches $d_{r}^{*}$, the required switching frequency in (\ref{freq_ZVS}) tends to decrease under PSPWM as shown in Fig. \ref{fig:fsw}. This reduction occurs when the inductor voltage nearly reaches zero, resulting in a very low volt-seconds. In other words, as $d^{*}$ nears $d_{r}^{*}$, the necessary switching frequency for achieving ZVS diminishes, making ZVS unfeasible. To address this limitation, the second/third-nearest voltage level is applied instead of the first/second-nearest one, using skipped-adjacency PWM (SAPWM). This method maintains an average pole voltage within switching period, and ZVS operation becomes feasible by modifying the pole voltage levels.

\begin{figure}[tbp]
    \centering
    \includegraphics[width=0.9\linewidth]{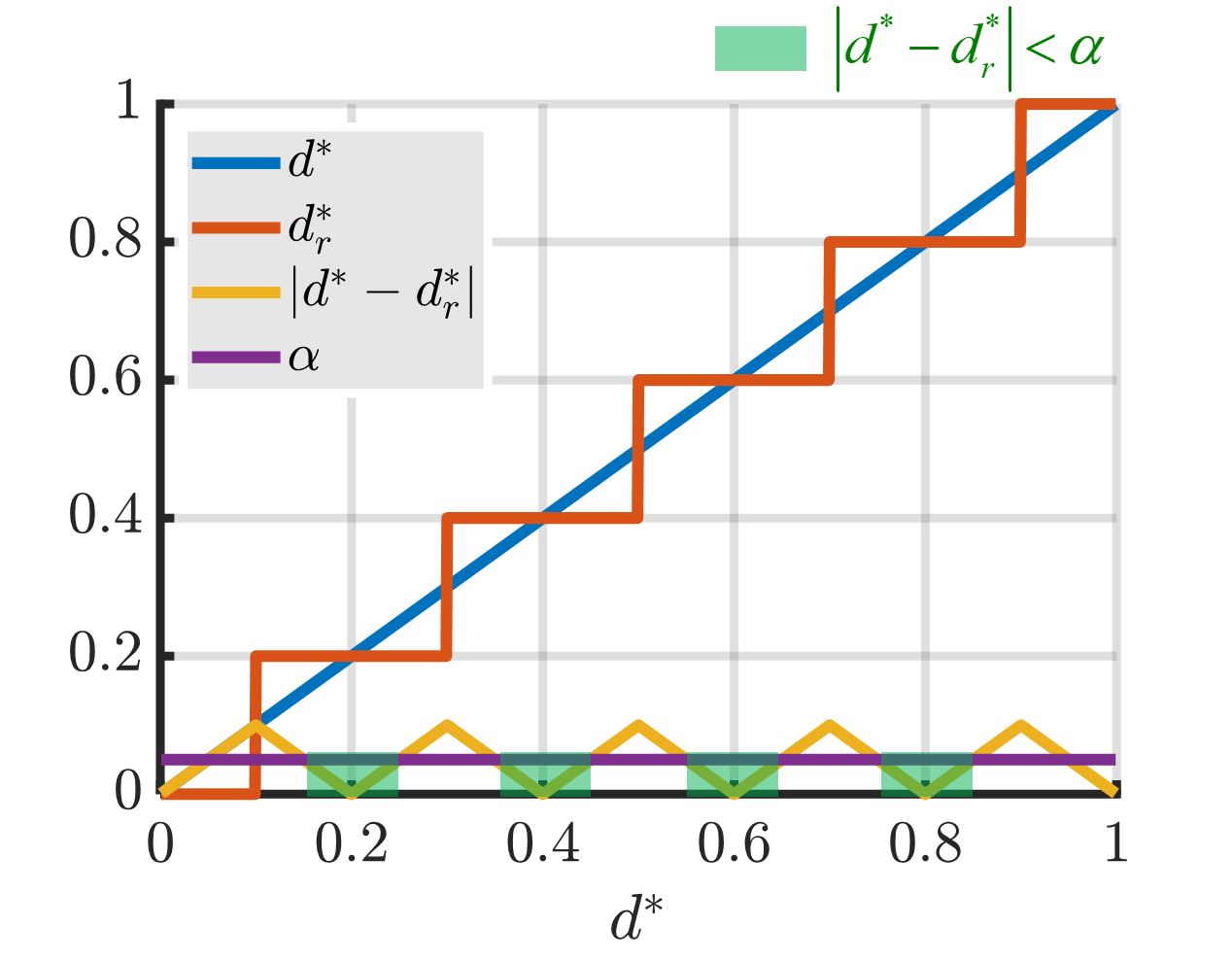} 
    \caption{Plot displaying the adjacency of the duty cycle to quantized duty levels, with a green region indicating the duty range requiring SAPWM.}
    \label{fig:duty_rel} 
\end{figure}
\begin{figure}[tp]
    \centering
    \includegraphics[width=0.9\linewidth]{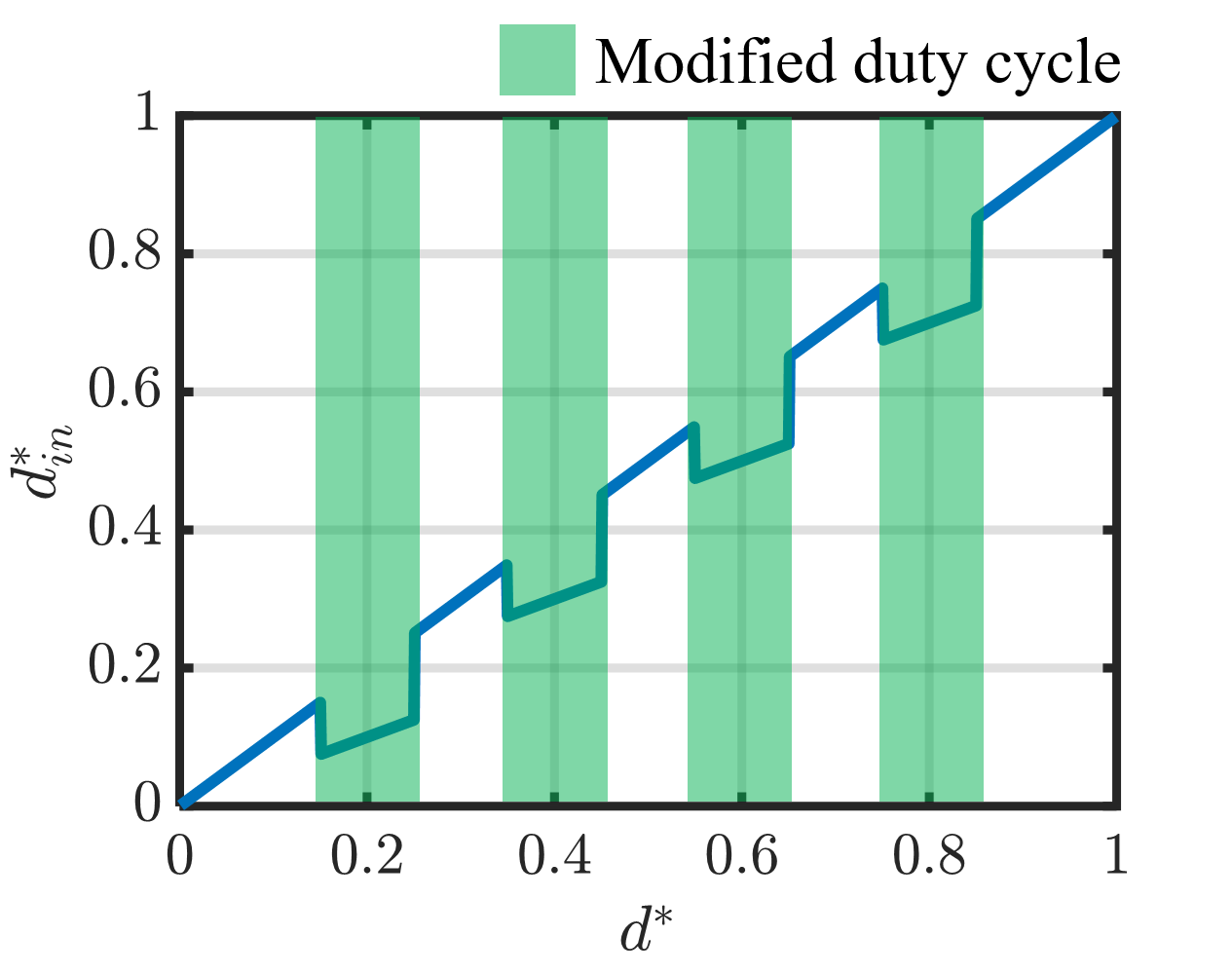} 
    \caption{$d^{*}_{in}$ graph showing modification of the duty cycle reference to achieve effective pole voltage in SAPWM.}
    \label{fig:dmod} 
\end{figure}
\begin{figure}[tp]
    \centering
    \includegraphics[width=0.9\linewidth]{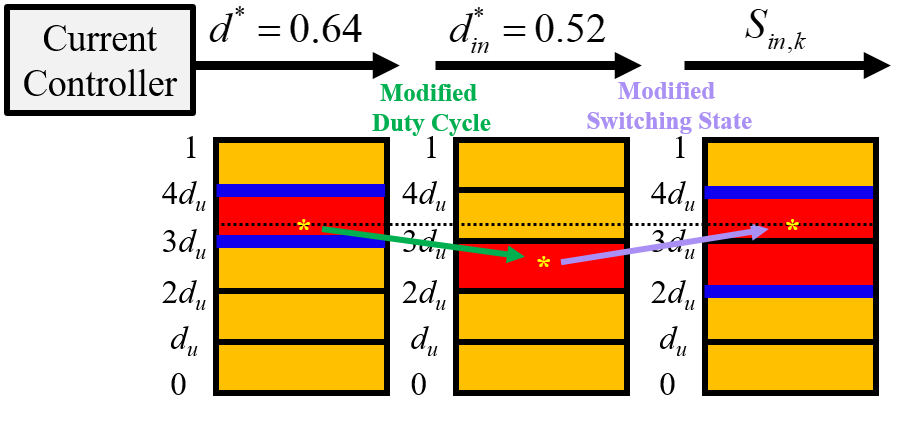} 
    \caption{Conceptual diagram showing how effective duty cycle is maintained in SAPWM through $d^{*}_{in}$  and $S_{in,k}$, when $N = 6$.}
    \label{fig:d_change} 
\end{figure}

\begin{figure}[tp]
    \centering
    \includegraphics[width=0.9\linewidth]{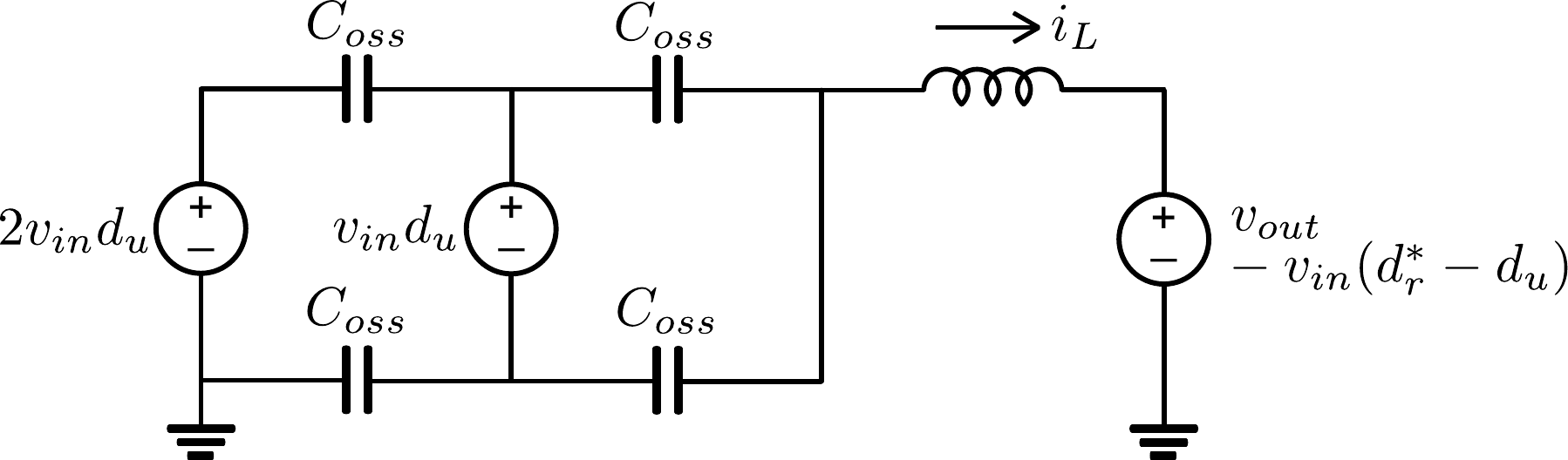} 
    \caption{Equivalent resonant circuit between the switch output capacitors and inductor under SAPWM during dead-time.}
    \label{fig:sa_res} 
\end{figure}
\begin{figure*}[tp]
    \centering
    \includegraphics[width=0.7\textwidth]{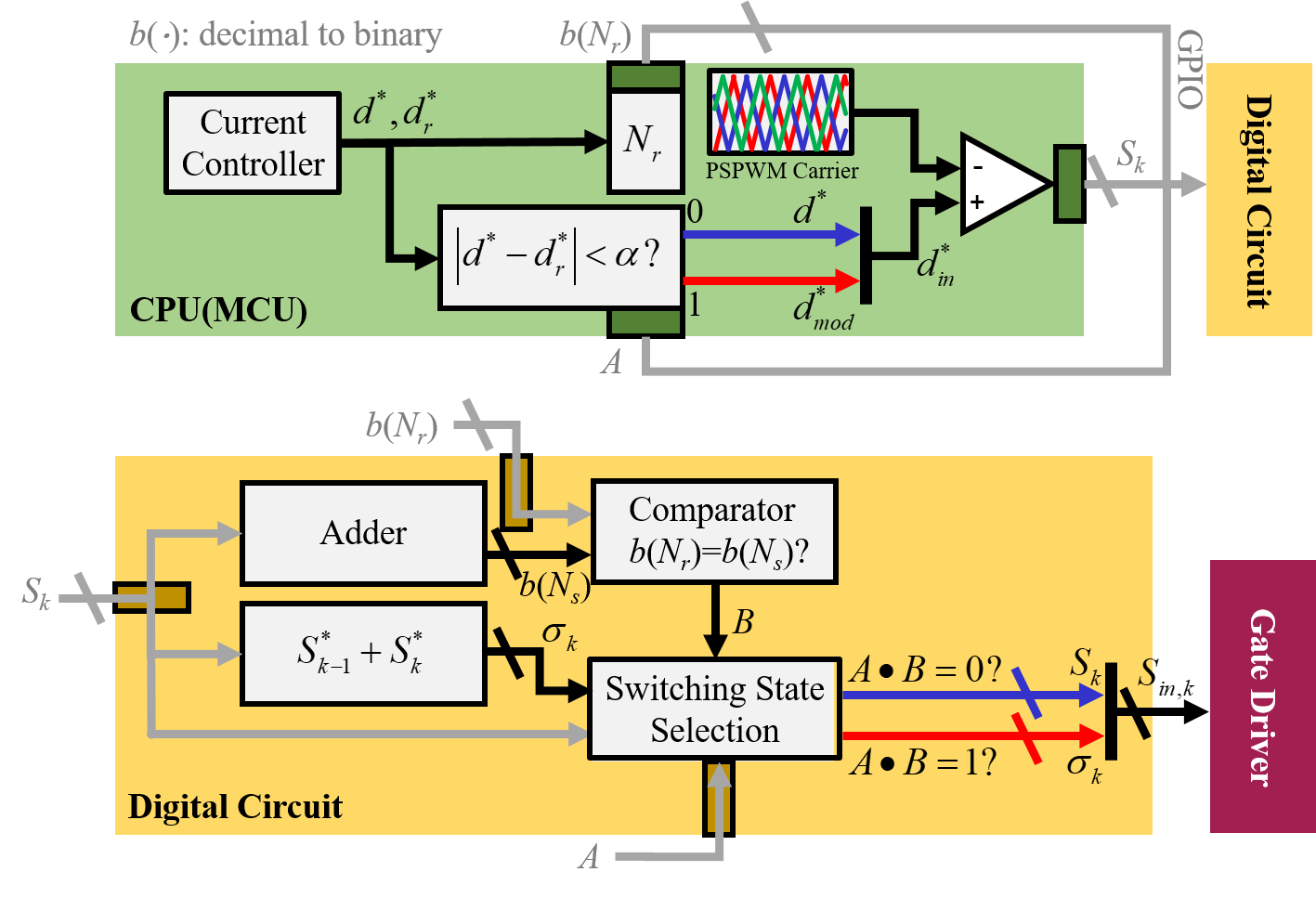} 
    \caption{Block diagram of the MCU and digital circuit, showing SAPWM implementation through duty cycle reference change in the MCU and logic operations in the digital circuit for switching state modification.}
    \label{fig:Control} 
\end{figure*}
\begin{figure*}[tp]
    \centering
    \includegraphics[width=1\textwidth]{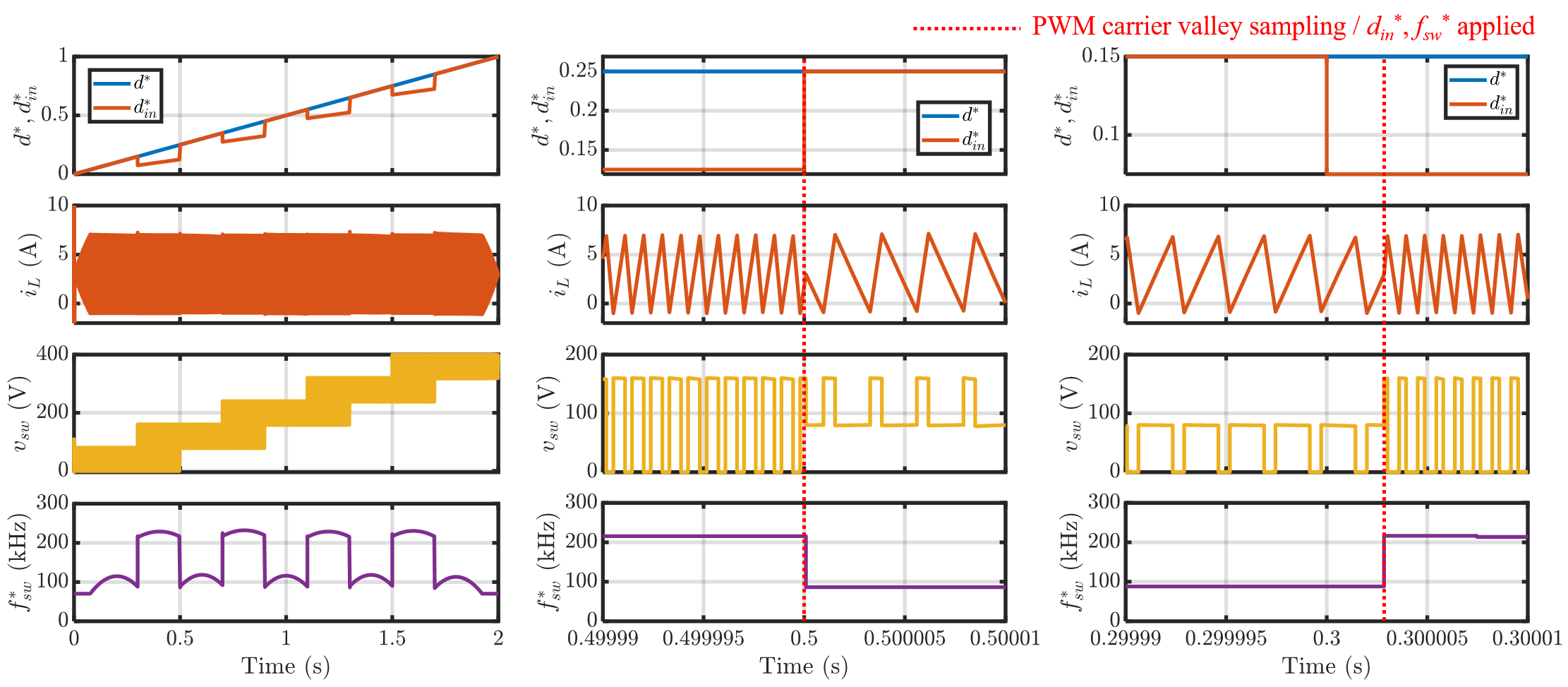} 
    \caption{ZVS current control over the full duty cycle range with SAPWM, substituting flying capacitors with independent voltage sources.}
    \label{fig:sim_ideal} 
\end{figure*}

To address regions where ZVS operation is infeasible under PSPWM due to excessively low switching frequency, the pole voltage of the most adjacent level is avoided. Following logic operation can be considered for this: 
\begin{equation}
{{S}_{k-1}}+ {{S}_{k}}={{\sigma }_{k}}
\label{mod_PWM}
\end{equation}
where, ${{S}_{N}}={{S}_{1}}$ and $k\in \left\{ 1,2,\cdots ,N-1 \right\}$. `+' is logical OR operator. 

Since each switch shares the same duty reference ($d^{*}$), the switching states of each switch under PSPWM align, causing adjacent switches to remain turned on or off together. Consequently, utilizing the OR operation between original switching state ($S_k$) and shifted switching state ($S_{k-1}$) in (\ref{mod_PWM}) raises the level of the resulting PWM pole voltage by one level, $\frac{v_{in}}{N-1}$. Thanks to the OR operation applied to adjacent switching states, each switching event either turns certain switches on or off in a straightforward manner. This means that there are no instances of simultaneous on-to-off and off-to-on transitions within the same event, which would otherwise lead to hard switching.

To avoid using the two closest pole voltage levels relative to the PWM voltage reference in PSPWM and instead apply the second and third closest pole voltage levels in SAPWM, each pole voltage level should be modified as follows:
\begin{equation}
\begin{aligned}
&\left[ {{v}_{in}}{{d}_{f}}^{*},{{v}_{in}}\left( {{d}_{f}}^{*}+{{d}_{u}} \right) \right] \\
&\to \left[ {{v}_{in}}\left( {{d}_{r}}^{*}-{{d}_{u}} \right),{{v}_{in}}\left( {{d}_{r}}^{*}+{{d}_{u}} \right) \right]
\end{aligned}
\end{equation}

To implement this, the modified switching state is set as follows:
\begin{equation}
S_{mod,k} = 
\begin{cases} 
\sigma_k & \text{if } N_r = N_s \\
S_k & \text{else}
\end{cases}
\end{equation}where ${{N}_{r}}=\left( N-1 \right){{d}_{r}}^{*}$ and ${{N}_{r}}=\sum\limits_{k=1}^{N-1}{{{S}_{k}}}$.

However, using the modified switching state ($S_{mod,k}$) results in a higher average PWM voltage than the original duty reference would produce due to the increased one of the pole voltage levels. To maintain an equivalent average PWM output voltage, the duty reference is modified as follows:
\begin{equation}
{{d}^{*}_{mod}}=\frac{{{d}^{*}}+{{d}_{r}}^{*}-{{d}_{u}}}{2}
\end{equation}

Finally, by using the adjacency threshold ($\alpha$), the regions of SAPWM and PSPWM can be divided as shown in Fig. \ref{fig:duty_rel}. The resulting duty reference ($d^{*}_{in}$), used as the input to the PWM comparator, and the final switching state ($S_{in,k}$), used as the input to the gate driver, are set as follows:
\begin{equation}
d_{in}^{*} = 
\begin{cases} 
      d^{*} & \text{if } \left| d^{*} - d_{r}^{*} \right| > \alpha \\
      d_{mod}^{*} & \text{else}
   \end{cases}
\end{equation}

\begin{equation}
S_{in,k} = 
\begin{cases} 
      S_{k} & \text{if } \left| d^{*} - d_{r}^{*} \right| > \alpha \\
      S_{mod,k} & \text{else}
   \end{cases}
\end{equation}

Although $d^{*}_{in}$ decreases in SAPWM region as in Fig. \ref{fig:dmod}, the switching state output from the MCU is post-processed by a digital circuit. This ensures that the average PWM voltage applied remains identical to the voltage reference. The conceptual diagram of the signal processing for SAPWM, shown in Fig. \ref{fig:d_change}, shows how the effective duty cycle remains consistent through $d^{*}_{in}$ and $S_{in,k}$.

When utilizing SAPWM, the applied current ripple nearly doubles, resulting in an increased switching frequency. Under SAPWM, the equivalent resonance circuit diagram during dead-time can be represented as shown in Fig. \ref{fig:sa_res}. Under PSPWM, the output capacitors of a single switch set are charged and discharged by the inductor as shown in Fig. \ref{fig:basic_res}. However, under SAPWM, ZVS requires that the output capacitors of two adjacent switch sets be charged and discharged by the inductor. Consequently, the amount of charge that must be transferred by $i_L$ has doubled compared to the previous requirement. Assuming the inductor energy is sufficiently large, the relationship between $I_{ZVS}$, $d_{t}$, and $C_{oss}$ should be met as follows:
\begin{equation}
\frac{4{{C}_{oss}}{{d}_{u}}{{v}_{in}}}{{{I}_{ZVS}}}\le {{t}_{d}}
\label{ineq2}
\end{equation}
Using the sampled current, the variable frequency reference under SAPWM can be calculated as follows:
\begin{equation}
    {{f}_{sw}}^{*}=\frac{\left\{ {{v}_{in}}\left( {{d}_{r}}^{*}-{{d}_{u}} \right)-{{v}_{out}} \right\}\left( {{d}^{*}_{mod }}-{{d}_{r}}^{*}+{{d}_{u}} \right)}{2L\left( \left| {{i}_{L}} \right|+{{I}_{ZVS}} \right)}
\end{equation}

When implementing the logic circuit using MCU digital outputs, it is essential to synchronize the entire operation using a clock to prevent unnecessary switching caused by delays in each digital logic stage. The clock signal is generated upon each change in the switching state, with a slight delay added to ensure all digital states reach a steady state before the logic operations proceed.
Without applying dead-time within the MCU, a delay can be added to the rising edges of the final gate driver input, $S_{in}$, to prevent short circuits.

The block diagram, including the MCU and digital circuit for SAPWM implementation, is shown in Fig. \ref{fig:Control}. $f^{*}_{sw}$ is generated in code and applied to register within the MCU, and the PWM carrier period ($1/f^{*}_{sw}$) update must be synchronized with ($d^{*}_{in}$) update. In Texas Instruments' TMS320283XX's CPU, the register value of ePWM peripheral  can be applied to the actual ePWM carrier only at the valley of the PWM carrier in shadowed mode. Accordingly, sampling, PWM carrier update, and duty cycle (PWM comparator input) update should occur at the valley of the PWM carrier.

\section{Results}
\begin{figure*}[tp]
    \centering
    \includegraphics[width=1\textwidth]{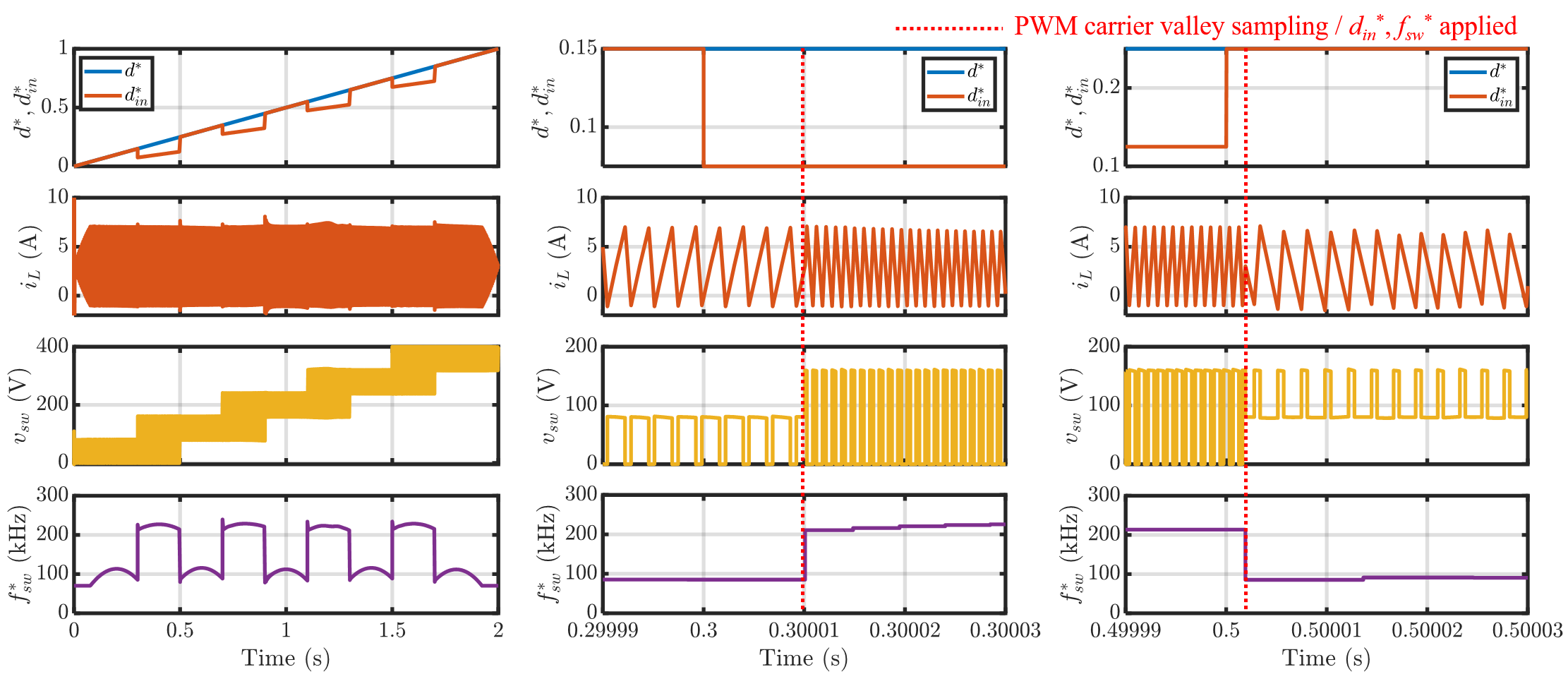} 
    \caption{ZVS current control over the full duty cycle range with SAPWM with 8.8 $\mu$F flying capacitors.}
    \label{fig:sim_real} 
\end{figure*}
\begin{figure}[tbp]
    \centering
    \includegraphics[width=0.8\linewidth]{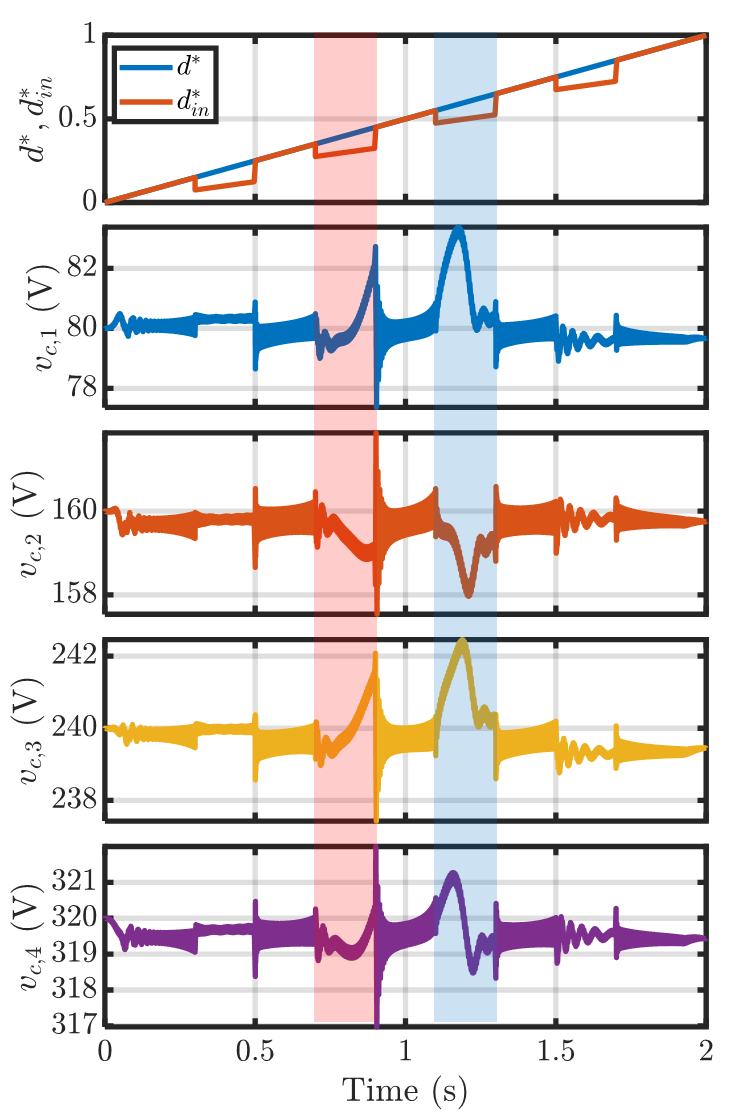} 
    \caption{Voltage variations in each flying capacitor during ZVS current control over the full duty cycle range using SAPWM.}
    \label{fig:vc} 
\end{figure}
To distinguish and compare the dynamics when using actual flying capacitors versus ideal voltage sources, the case of ideal voltage sources is examined at first. In this scenario, it is assumed that the flying capacitors are sufficiently large, allowing to ignore the dynamics of the flying capacitors.
\begin{table}[h!]
    \centering
    \begin{tabular}{|c|c|c|}
        \hline
        \textbf{Parameter} & \textbf{Notation} & \textbf{Value} \\
        \hline
        FCML Level & $N$ & 6 \\
        \hline
        Capacitance of flying capacitor & $C_f$ & 8.8 $\mu$F \\
        \hline
        Inductance of inductor & $L$ & 4.4 $\mu$H \\
        \hline
        Input voltage & $V_{in}$ & 400 V \\
        \hline
        \makecell{PWM carrier frequency \\ (switching frequency)} & $f_{sw}$ & 70-230 kHz \\
        \hline
        Effective switching frequency & $f_{sw,eff}$ & 350-1150 kHz \\
        \hline
    \end{tabular}
    \caption{Parameters for FCML Converter}
    \label{parameters}
\end{table}

Fig. \ref{fig:sim_ideal} shows the ZVS operation results when ideal voltage sources are utilized instead of flying capacitors. It confirms that full-range ZVS can be achieved within the duty cycle range of $\alpha < d^{*} < 1 - \alpha$ with the proposed SAPWM. In the regions where SAPWM is applied, the second and third nearest pole voltage levels are utilized, and the first nearest pole voltage is avoided. Moreover, the switching frequency required for ZVS increases with the volt-seconds rise. \\
\indent Since $d^{*}_{in}$ and $f^{*}_{sw}$ updates are synchronized with the PWM carrier valley during each PWM transition, a smooth and immediate transition occurs without any transients. Although $d^{*}_{in}$ decreases in the SAPWM region, subsequent logic operations on the  switching states from MCU maintain the applied average PWM voltage consistent with the pole voltage reference from the current controller.\\
\indent Fig. \ref{fig:sim_real} shows the results when using flying capacitors with the specifications listed in TABLE \ref{parameters}. Only passive balancing is used to achieve flying capacitor voltage balancing, without active balancing. In this case, ZVS operation is also achievable in a range of $\alpha < d^{*} < 1 - \alpha$, and the increased volt-seconds in the SAPWM region is observed. During transition between PSPWM and SAPWM regions, the dynamics of passive balancing change slightly, causing minor oscillations. In addition, stable operation of passive balancing is maintained even under SAPWM, allowing the entire system to converge to a steady state where the flying capacitors remain balanced. Since $d^{*}$ is set based on the input voltage without accounting for individual fluctuation in flying capacitor voltages, these variations shown in Fig. \ref{fig:vc} may slightly shift the peak and valley points of $i_{L}$, which could induce instantaneous hard switching. This issue can be addressed by designing the ZVS current with additional margin or by directly measuring the flying capacitor voltages and adjusting the duty cycle reference for each switch with active balancing \cite{10601504}. As a result, the simulation results demonstrate that full-range ZVS can be achieved with SAPWM with a small variation of peak and valley current due to flying capacitor voltage fluctuation under passive balancing.

\section{Future Work}
In Fig. \ref{fig:vc}, when SAPWM is applied around duty cycles of 0.4 and 0.6, passive balancing shows reduced bandwidth and underdamped behavior compared to PSPWM. This damping characteristics can be examined by comparing the dynamics of flying capacitors under SAPWM and PSPWM through state-space analysis. Furthermore, SAPWM in this range can be implemented with different digital logic configurations beyond the proposed method, each affecting passive balancing characteristics differently. 

Future work will focus on validating the proposed approach with hardware experiments, refining SAPWM logic to enhance the dynamic response of passive balancing, and conducting a mathematical analysis of passive balancing stability \cite{9829991, 8013368,6030953,4453867}.

\section{Conclusion}
The proposed SAPWM method broadens the ZVS operating range, overcoming limitations in achieving ZVS in a full duty cycle range with variable switching frequency under conventional PSPWM. By adjusting the duty reference for the PWM comparator input and applying logic operations to the switching state, the effective pole voltage is consistently maintained, while volt-seconds increase to enable full-range ZVS operation. Current control results confirm full-range ZVS, with only minor oscillations in the flying capacitor voltage observed during PWM mode transitions. These oscillations have a slight influence on inductor current peaks and valleys, indicating that such variations should remain within the ZVS range for reliable operation.
\ifCLASSOPTIONcaptionsoff
  \newpage
\fi

\bibliographystyle{IEEEtran} 
\bibliography{IEEEabrv}

\end{document}